\documentclass{article}
\usepackage[utf8]{inputenc}
\usepackage{amsmath}
\usepackage{siunitx}
\usepackage[autostyle]{csquotes}
\MakeOuterQuote{"}
\usepackage{subcaption}
\usepackage{float}
\usepackage{physics} 
\usepackage{xcolor} 
\usepackage{upgreek}
\usepackage[demo,abs]{overpic}
\usepackage{multirow}
\usepackage{authblk}

\title{Lasing modes in ZnO nanowires coupled to planar metals}
\author[1]{Daniel Repp}
\author[1,2]{Angela Barreda}
\author[2]{Francesco Vitale}
\author[1,2]{Isabelle Staude}
\author[3]{Ulf Peschel}
\author[2]{Carsten Ronning}
\author[1,4]{Thomas Pertsch}
\affil[1]{Institute of Applied Physics, Abbe Center of Photonics, Friedrich Schiller University Jena, Albert-Einstein-Straße 15, 07745 Jena, Germany}
\affil[2]{Institute of Solid State Physics, Friedrich-Schiller-Universität Jena, Max-Wien-Platz 1, 07743 Jena, Germany}
\affil[3]{Institute of Condensed Matter Theory and Solid State Optics, Friedrich-Schiller-Universität, Max-Wien-Platz 1, Jena 07743, Germany}
\affil[4]{Fraunhofer Institute for Applied Optics and Precision Engineering, Albert-Einstein-Straße 7, 07745 Jena, Germany}
\date{}

\usepackage[
backend=biber,
style=numeric,
citestyle=numeric,
sorting=none,
doi=false,isbn=false,url=false,eprint=false
]{biblatex}

\renewbibmacro{in:}{%
  \ifentrytype{article}{}{\printtext{\bibstring{in}\intitlepunct}}}

\AtEveryBibitem{\clearfield{month}}
\AtEveryBibitem{\clearfield{pages}} 
\usepackage{graphicx}
\begin{document}

\maketitle

\begin{abstract}
Semiconductor nanowire lasers can be subject to modifications of their lasing threshold resulting from a variation of their environment. A promising choice is to use metallic substrates to gain access to low-volume Surface-Plasmon-Polariton (SPP) modes. We introduce a simple, yet quantitatively precise model that can serve to describe mode competition in nanowire lasers on metallic substrates. We show that an aluminum substrate can decrease the lasing threshold for ZnO nanowire lasers while for a silver substrate, the threshold increases compared with a dielectric substrate. Generalizing from these findings, we make predictions describing the interaction between planar metals and semiconductor nanowires, which allow to guide future improvements of highly-integrated laser sources.
\end{abstract}

\section{Introduction} 

\begin{figure}
  \centering
  \includegraphics[width=0.8\linewidth]{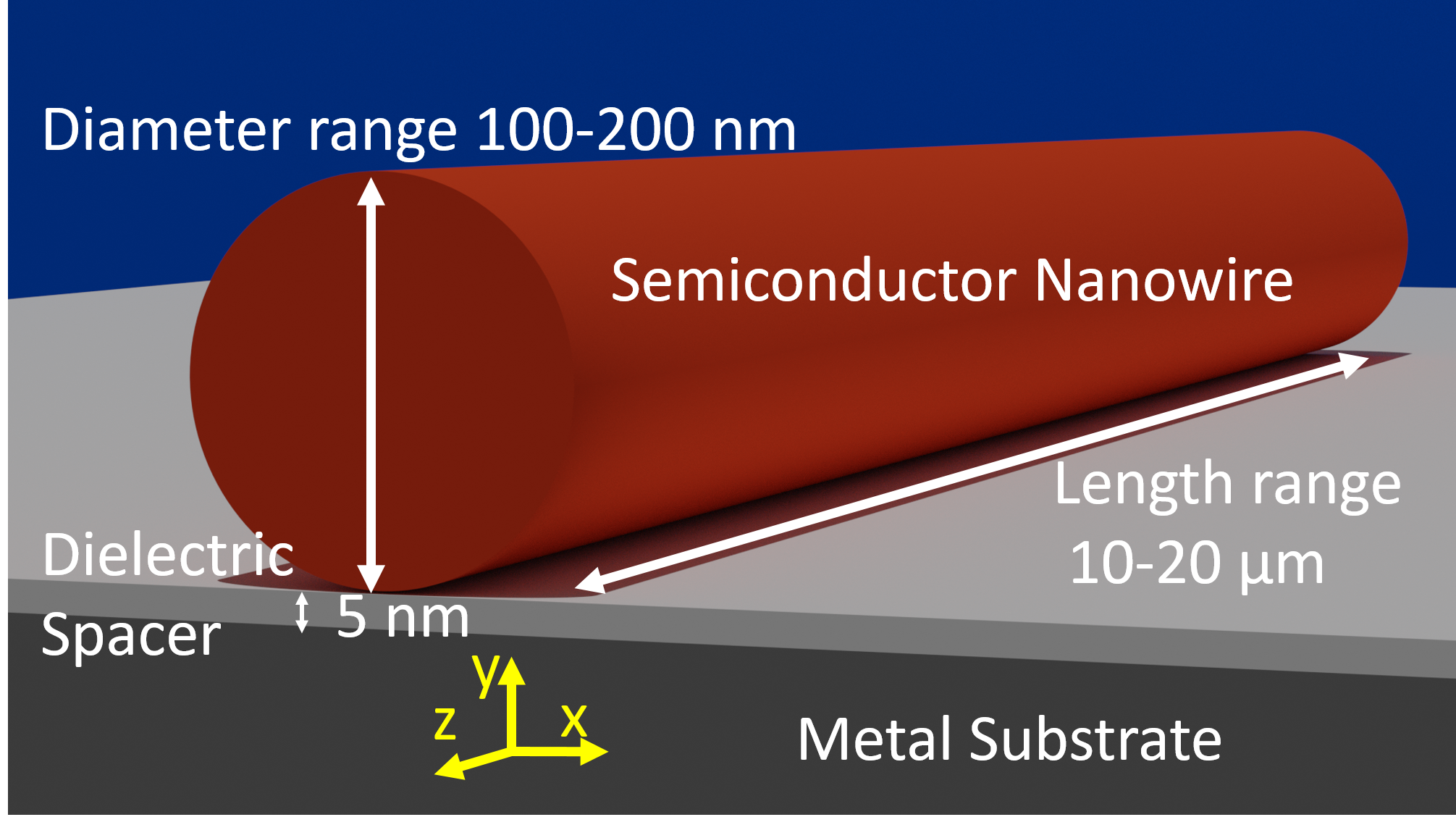}
\caption{Scheme of the system under investigation in this article (not to scale). The coordinate system used throughout this article as well as typical geometrical scales are indicated.}
\label{fig:wire_blender}
\end{figure}

Semiconductor nanowires have attracted a broad interest in recent years and have been studied in fields as diverse as nanosensing \cite{ZnOLaser}, nanoscale optoelectronics \cite{FET1} and thermoelectrics \cite{NanowireThermo}. They are of particular interest in the field of nanophotonics, as they are wavelength-scale nanolasers \cite{Nanolaser2020Review, Oulton2009}. ZnO nanowires are interesting for UV-applications due to their direct bandgap in this spectral range, corresponding to an emission wavelength $\lambda \approx \SI{380}{\nano\meter}$, and high exciton binding energy, supporting excitonic emission and electron-hole-plasma formation even at room temperature \cite{ZnOReview}. \ \\
Nanowire-based lasers form a cylindrical, quasi-one-dimensional cavity and have transversal geometrical extents close to or below optical wavelengths. Since the fabrication of Bragg-reflectors during nanowire growth is challenging at the nanoscale \cite{feedback}, the cavity feedback is often restricted to the back-reflection of the end facets of the nanowires. Furthermore, the lasing properties of nanolasers can be changed by interaction with the substrate. For example, the radiation emitted by the nanowire can interact with the free electron plasma of metallic substrates, forming SPPs \cite{InGas1, InGas2}, which can facilitate a reduction of the mode volume \cite{StimulatedSPP2}. Metals known to provide this mode confinement for ZnO nanowires have plasma frequencies in the ultraviolet, like aluminum and silver \cite{AluminumPlasmonics}.\ \\
The reduced mode volume ($V$) leads to a faster emission dynamics via the Purcell effect \cite{SubwavelengthDynamicsPurcell}, which is quantitatively described by the Purcell factor $F$ \cite{SVG}:
\begin{equation}
    F = \frac{\gamma}{\gamma_0} = \frac{3\lambda^3}{4\pi^2} \frac{Q}{V},
    \label{PurcellDefinition}
\end{equation}
where $\gamma$ is the emission rate of a recombining electron-hole pair in the structure of interest (here the nanowire on top of the metal), $\gamma_0$ is the emission rate of the same electron-hole pair in bulk ZnO, $\lambda$ is the emission wavelength, $Q$ is the quality factor of the resonance and $V$ is the mode volume.
Usually, a small dielectric spacer is sandwiched between the nanowire and the underlying metal substrate to protect the nanowire emission from quenching by the metal \cite{SpacerQuenching}. An illustration of the entire structure can be seen in Fig. \ref{fig:wire_blender}.

The Purcell effect, which is applicable to emission of radiation in general, is discussed in many different lasing applications \cite{SemLaser, PurcellPC}. In fact, the system described above (nanowire on a metallic substrate) is often called a spaser, which is an acronym for "Surface Plasmon Amplification via Stimulated Emission of Radiaton", in analogy to the laser \cite{SpaserDef}. The physical principle of a spaser is equivalent to that of a laser, only the amplified bosonic mode is a surface plasmon polariton instead of a photon \cite{StimulatedSPP}. \ \\
In this article, a thorough investigation of the influence of nanowire size and substrate material on the Purcell factor is presented. This information is combined with detailed mode calculations and used in a lasing model to calculate mode-specific lasing thresholds and thereby quantify mode competition in the system. As a result, we gain fundamental, non-trivial insights into the mode competition processes for ZnO nanowire lasers coupled to planar metals and highlight trade-offs that have to be considered in the device design. Furthermore, we provide a versatile, quantitative tool for the description, investigation and optimization of semiconductor lasers in the spaser geometry. Also, we define two coupling regimes between SPPs and nanowire emission, having distinct advantages and disadvantages.

\section{Numerical methods}
\label{Simulation_configuration}

\subsection{Mode calculation}
\label{Confined_modes}An important part of the quantitative description of the lasing process is the proper characterization of the lasing modes regarding critical parameters like propagation losses and gain overlap. The MODE package of the Lumerical software suite was used in order to obtain the transversal eigenmodes of the simulated structures \cite{Lumerical}. This package implements an EigenMode Expansion (EME) on a Yee-grid. The rectangular Yee grid is known to lose accuracy for metal interfaces and curved surfaces, both of which are present in our system. Treating the system therefore often requires very fine grids. In order to obtain highly accurate results, a technique called "Conformal Meshing" is used, which calculates effective permittivities close to interfaces and greatly reduces the computational costs.\ \\
To correctly connect the Purcell factor and the different mode profiles, a normalization procedure of the modes has to be chosen. The Purcell factor describes the modification of emission rates as a result of changes in the environment. In quantum mechanics, Fermi's Golden rule describes the rate of decay and, correspondingly, photon emission. The matrix element responsible for the description of decay process in Fermi's Golden rule, however, is proportional to the field amplitude. Therefore, a normalization with respect to the field strength has to be chosen, which is defined as

\begin{equation}
    \int_A \sum_{j}|E_{i,j}(\textbf{r})|^2 = 1,
\end{equation}

where $A$ is a cross section through the mode perpendicular to the nanowire axis and $E_{i,j}(\textbf{r})$ is the $j$ component of the electric field strength of mode $i$ at position $\textbf{r}$. 

\begin{figure}
 \includegraphics[width=1\linewidth]{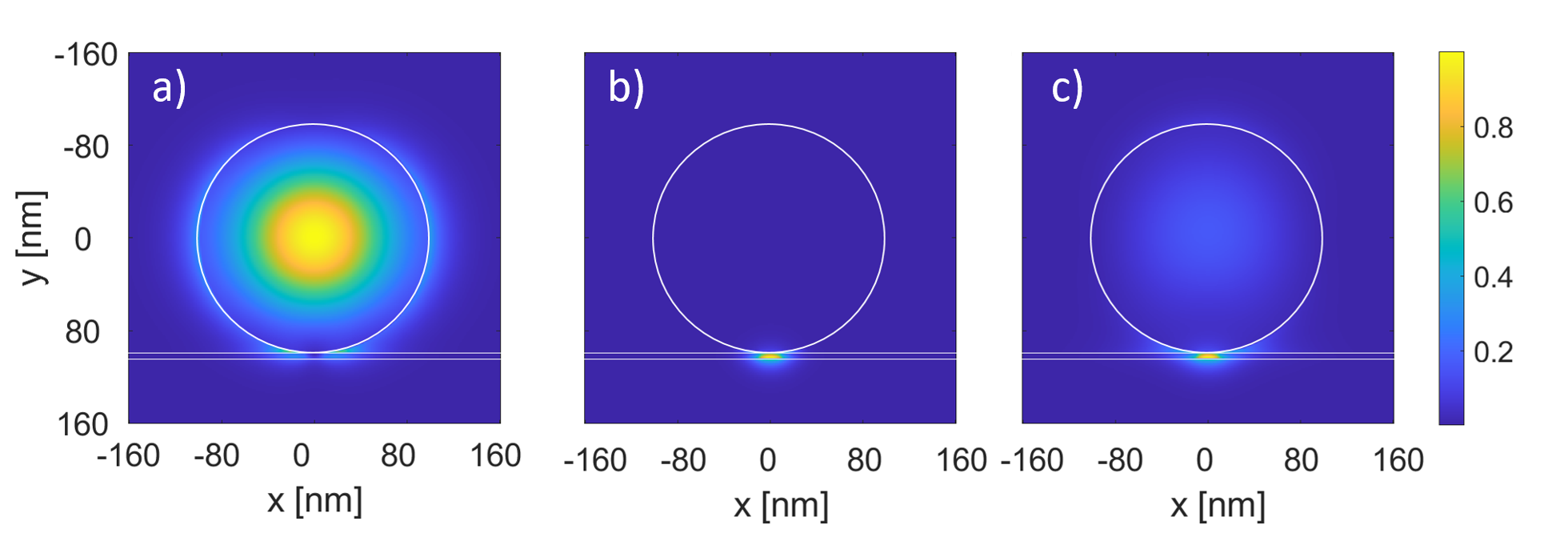}
\caption{Example plots of $|E|^2$ of the considered modes for a ZnO nanowire of a diameter of $\SI{200}{\nano\metre}$ on top of a silver substrate with an interposed silica nanometric spacer. Plotted are the a) photonic mode ($n_{eff} = 2.1+0.002i$), b) plasmonic mode ($n_{eff} = 3.78+1.76i$) and c) hybrid mode ($n_{eff} = 2.07+0.02i$).}
\label{Modes}
\end{figure}
The profiles of some of the transversal modes are plotted in Fig. \ref{Modes}. They are similar to the ones described in ref. \cite{Modes}. The plots include the nanowire, the metallic substrate and a dielectric spacer in between. We call the nanowire bound mode with a strong maximum in the nanowire center, and almost no visible interaction with the metal, a "photonic mode", as plotted in Fig. \ref{Modes}a.\ \\
The main interest of this investigation is however on highly confined "plasmonic" modes. These modes exhibit a very small mode volume which leads to a large Purcell factor. The fields of plasmonic modes are mainly confined within the spacer, as shown in Fig. \ref{Modes}b.\ \\
A third "hybrid" mode can consists of a nanowire waveguide mode, which is dominantly polarized orthogonally to the metallic substrate. Compared to the plasmonic mode, the overlap with the nanowire material is increased. An example is plotted in Fig. \ref{Modes}c.

\subsection{Purcell factor determination}
\label{NumericalMethods_Purcell}
The Purcell factor was determined by performing Finite-Difference-Time-Domain (FDTD) simulations \cite{Lumerical}, utilizing a dipole source placed at points inside the nanowire, which maximize the coupling to the modes described in Sec. \ref{Confined_modes}. To characterize the photonic mode, the dipole was placed at the nanowire center. For the plasmonic- and hybrid modes, the dipole was placed closely to the metal. Its emission wavelength was set to $\lambda = \SI{380}{\nano\metre}$, which corresponds to the bandgap of ZnO. \ \\
The \textbf{E} and \textbf{H} fields emitted by the dipole were recorded on the six sides of a small cube (side length of 5 nm) placed around the emitter. The Fourier-transforms of the recorded fields were subsequently used to obtain the spectrum of Poynting vectors, which was integrated over all sides of the cube, yielding the spectral power distribution \cite{LumericalPoynting}:

\begin{equation}
    P(\omega) = \frac{1}{2} \int_{\text{Box}} Re(\textbf{E}(\omega)\times\textbf{H}^*(\omega)) dA.
\end{equation}

\noindent This distribution was subsequently normalized to the injected source power, yielding the Purcell factors $F_j$ by arguments presented in the supplementary document Sec. 1A, where $j$ indicates the dipole orientation.\ \\
To attribute the Purcel factor calculated at position $\textbf{r}$ to a specific mode $i$, the normalized mode profiles $\textbf{E}_i(\textbf{r})$ are used. At position $r$ and with orientation $j$, the Purcell factor attributed to mode $i$ was calculated as

\begin{equation}
    F_{i,j}(\textbf{r}) = F_j(\textbf{r}) \cdot \frac{|E_{i,j}(\textbf{r})|^2}{\Sigma_l |E_{l,j}(\textbf{r})|^2},
\end{equation}

where the index $l$ runs over all modes.\ \\
The Purcell factor, as defined above, is a vectorial quantity. To describe the Purcell factor attributable to mode $i$ as a scalar value in a rate-equation model, the stochastic distribution of dipole orientations has to be taken into account. Assuming the electron-hole plasma responsible for gain in ZnO nanowire lasers to be isotropic, the dipole orientation was considered equally distributed, leading to a simple averaging procedure \cite{AverageF}:

\begin{equation}
    F_i(\textbf{r}) = \frac{\sum_{j}F_{i,j}(\textbf{r})}{3}
\end{equation}

where $j$ indicates the dipole orientation.\ \\
The nanowire was modeled as an isotropic, non-dispersive dielectric material with a refractive index of $2.5$ to conform to the conditions at the emission wavelength of $\lambda = \SI{380}{\nano\metre}$.  The chosen refractive index value corresponds approximately to the value of the refractive index of crystalline ZnO from ref. \cite{ZnO_Model} and is supported by a calculation from ref. \cite{VersteeghScreening} at the Mott density $N_{\text{Mott}} = 2\cdot 10^{24} \si{\per\metre\cubed}$ (value from \cite{VersteeghCLEO}), as can be seen also in ref. \cite{ChargeCarriers}. The actual value is seen to vary within a 10-20\% range, which is an effect that was neglected to keep the model simple. The possible errors caused by this particular choice are elaborated on in Sec. \ref{ThresholdTheory} and in the supplementary document Sec. 2D. The imaginary part of the refractive index was set to zero, as losses and gain are incorporated into the gain profile in the lasing model (see Sec. \ref{LasingModel}).\ \\
The nanowire was placed on top of a silica spacer-layer with a thickness of 5 nm, which separates the nanowire from a 100 nm thick metal layer, with the metal being either aluminum or silver. The simulation domain was discretized with a uniform mesh of 1 nm to guarantee the convergence of the results. Perfectly Matched Layer (PML) boundary conditions were applied to avoid reflections on the domain boundaries. Both nanowire end facets as well as the metal were extended into the PML to disable end facet reflections, which effectively corresponds to an infinitely long nanowire. Reflections at the nanowire end facets were addressed with a separate FDTD calculation (see supplementary document). \ \\

\subsection{Lasing model}
\label{LasingModel}
A multimode rate-equation lasing model, adapted from ref. \cite{LasingModel}, was used to describes the lasing process. Here, a pumping rate $P$ describes optical excitation of charge carriers. After the relaxation of charge carriers, they recombine and emit photons, which couple to specific modes $i$ as described by the Purcell factors $F_i$ calculated above. The rate equations governing the evolution of the charge carrier density $n$ as well as the photon densities $s_{i}$ in the three competing modes as shown in Fig. \ref{Modes} are:

\begin{align}
    \dv{n}{t} &= \eta \frac{P}{\hbar\omega V_\text{a}} - \frac{n}{\tau_{\text{nr}}} - (F_1\beta_1 + F_2\beta_2 + F_3\beta_3) \frac{n}{\tau_{\text{sp}}}  \nonumber\\
    &\quad -\nu_{\text{g,a}} g(n) (\Gamma_1 F_1 s_1 + \Gamma_2 F_2 s_2 + \Gamma_3 F_3 s_3) \label{LasingModelCarrierDensity} \\
    \dv{s_1}{t} &= F_1\beta_1 \frac{n}{\tau_{\text{sp}}} + \nu_{\text{g,a}} g(n) \Gamma_1 F_1 s_1 -\frac{s_1}{\tau_{\text{p,1}}} \label{LasingModelMode1} \\
    \dv{s_2}{t} &= F_2\beta_2 \frac{n}{\tau_{\text{sp}}} + \nu_{\text{g,a}} g(n) \Gamma_2 F_2 s_2 -\frac{s_2}{\tau_{\text{p,2}}} \\
    \dv{s_3}{t} &= F_3\beta_3 \frac{n}{\tau_{\text{sp}}} + \nu_{\text{g,a}} g(n) \Gamma_3 F_3 s_3 -\frac{s_3}{\tau_{\text{p,3}}} .
\end{align} 

In this model, $\eta$ is the absorption efficiency of the optical pump, which is set to 1 to model efficient one-photon above-bandgap excitation, $P$ is the pumping rate, $\hbar$ is the reduced Planck constant, $\omega$ is the incoming photon angular frequency and $V_\text{a}$ is the volume of the active medium. $1/\tau_{\text{nr}}$ is the nonradiative decay rate. This rate is kept constant, which neglects higher order nonradiative recombination processes like Auger recombination, to simplify the model. $1/\tau_{\text{sp}}$ is the rate of spontaneous emission. The exciton lifetime $\tau_{total}$ in ZnO nanowire is in the range of 200-300 ps \cite{Exciton_Lifetime_1, Exciton_Lifetime_2}, combining radiative ($\tau_{sp}$) and non-radiative ($\tau_{nr}$) lifetime. Measurements of the Internal Quantum Efficiency (IQE) yield values close to 50\% \cite{IQE}. Therefore, we assume that $\tau_{sp} = \tau_{nr}$, resulting in

\begin{align}
    \frac{1}{\tau_{total}} &= \frac{1}{\tau_{sp}} + \frac{1}{\tau_{nr}}  \\
    &= \frac{2}{\tau_{sp}}, 
\end{align}

which yields $\tau_{sp} = \tau_{nr} = 2\tau_{total} = \SI{700}{\pico\second}$. The confinement factors $\Gamma_{i}$ were defined as

\begin{align}
    \Gamma_i &= \frac{\int_{V_\text{a}}\epsilon(\textbf{r})|E_i(\textbf{r})|^2}{\int_{V}\epsilon(\textbf{r})|E_i(\textbf{r})|^2} \label{confinementV}\\
    &\approx \frac{\int_{A_\text{a}}\epsilon(\textbf{r})|E_i(\textbf{r})|^2}{\int_{A}\epsilon(\textbf{r})|E_i(\textbf{r})|^2}.
    \label{confinementA}
\end{align}

This expression, although not accurate for open systems, leads to reliable results. Special care is necessary for open systems, which can be treated by quasi-normal modes \cite{QNM}. In Eq. \ref{confinementV}, $V$ is the total volume of the simulation domain, $\epsilon(\textbf{r})$ is the permittivity distribution and $\textbf{E}_i(\textbf{r})$ the field profile of mode $i$. The integration along the nanowire axis cancels out because of translation symmetry, leaving the actual integration to be performed on the nanowire cross section $A_a$ and the cross section of the simulation domain $A$. There is a slight error caused by the termination of the nanowire which decreases with increasing nanowire length.\ \\
Note that, while $\Gamma_i$ is calculated for the entire simulation domain, the value is dominated by the maximum of the specific mode. Even more so because of the high value of the nanowire permittivity index. This justifies the connection of a confinement factor $\Gamma_i$ and a Purcell factor $F_i$ calculated for only one dipole position.\ \\ 
Also, note that the fraction of spontaneous emission into mode $i$ was chosen as $\beta_i = \frac{\Gamma_i}{\Gamma_1+\Gamma_2+\Gamma_3}$. This choice is a sum over all longitudinal cavity modes in the nanowire having the same transversal mode profile. It neglects all other modes except for the ones considered, as well as emission in free space, and weights the amount of spontaneous emission into mode $i$ by its overlap with the gain material. Note that this is a heuristic approach motivated by ref. \cite{BetaFDTD}.\ \\
Furthermore, the group velocity is defined as

\begin{equation}
    \nu_{\text{g,a}} = \frac{c}{\dv{(\omega n')}{\omega}},
\end{equation}
with $n'$ being the active medium refractive index, and $g(n)$ is the active medium gain as in \cite{LasingModel}, given by

\begin{equation}
    g(n) = g_0\ln(\frac{n}{n_{\text{tr}}}),
\end{equation}

with $n_{\text{tr}} = n_{\text{mott}} = 2\cdot 10^{24} \si{\meter\cubed}$ the transparency density, which we set equal to the Mott density for formation of an electron-hole plasma \cite{VersteeghScreening}. The transparency density changes with wavelength, but the Mott density gives a good order-of-magnitude estimate. Furthermore, we set $g_\text{0}=\SI{0.7}{\per\micro\meter}$ \cite{gainRef}. \ \\
The modal losses are modelled via a loss rate, given by

\begin{equation}
    \frac{1}{\tau_{\text{P},i}} = \nu_{\text{gz},i} \left(\alpha_i + \frac{1}{2L} \ln(\frac{1}{R_i^2}), \right)
\end{equation}
where $\nu_{\text{gz},i} = c/n'_i$, with $n'_i$ being the mode group index, is the mode group velocity, $\alpha_i$ is the modal propagation loss, $L$ the length of the nanowire (we set $L=\SI{10}{\micro\metre}$) and $R_i$ the end facet reflectivity for mode $i$.\ \\

\subsection{Thresholds}
\label{ThresholdTheory}
To identify the lasing threshold, the initial value problem defined by the lasing model was solved using an implicit multi-step Runge-Kutta time-stepping algorithm, implemented in the Python SciPy package. The initial charge carrier density was set to $1 \si{\per\meter\cubed}$ to treat an ideal, defect-free semiconductor without intrinsic doping while avoiding a diverging logarithm. The initial photon densities were set to  $\SI{0}{\per\meter\cubed}$. The simulation time was set to $\SI{10}{\nano\second}$, but the results do not depend on the simulation time in a window of $\SI{100}{\pico\second}$ to $\SI{10}{\nano\second}$. \ \\
Sweeps of the pumping rate $P$ were performed until characteristic relaxation oscillations of the photon number are observed, which are identified by comparing the maximum and steady-state values of the photon numbers in the time trace. If they differed by more than 10\%, the corresponding power is defined as the threshold power, and subsequently divided by the area of a typical, circular laser spot size of $\SI{10}{\micro\meter}$ to obtain the intensity.\ \\
The nanowire material model has an impact on the accuracy of the threshold calculation, since the chosen refractive index is on the upper end of the likely values and changes during the build-up of the electron-hole plasma. First of all, it overestimates the end facet reflectivity, leading to an underestimation of the thresholds, particularly of the photonic modes. On the other hand, it also underestimates the Purcell factor of plasmonic/hybrid modes, leading to an overestimation of the threshold. These effects cause a correction of the results, but can be  incorporated into the model by a careful, piecewise modeling of the loss channels. Detailed plots showing these error sources can be found in the supplementary document.\ \\

\subsection{FDTD lasing calculations}
\label{FDTD}
Additionally to the model presented in Sec. \ref{LasingModel}, full-wave FDTD calculations were performed for selected nanowire diameters in order to corroborate the developed approximative lasing model. \ \\
The gain model used was the 4-level-2-electron material with a pumping level at 350 nm (FWHM 0.8 nm) and a broad lasing level at 380 nm (FWHM 8 nm). The nanowire was pumped with a Total-Field-Scattered-Field (TFSF) source \cite{Lumerical}, which is essentially a uniform source, encompassing the whole nanowire, with an amplitude of $5\cdot 10^7\cdot \si{\volt\per\meter}$ for an aluminum and $10^8\cdot \si{\volt\per\meter}$ for a silver substrate and a wavelength resonant with the pumping transition of the 4-level system. The pump was polarized in the direction of the nanowire axis and the duration was set to 500 fs for the aluminum and 1000 fs for the silver substrate.\ \\
The nanowire end facets were not extended into the PMLs. The nanowire length was chosen as 3 µm. A non-uniform mesh was used to decrease computational requirements because of the large simulation area and the simulation duration.\ \\
When the pump is shut off, broadband temporal Fourier-components are excited, some of which are resonant with the lasing wavelength, seeding the emission of a pulse. The spatial and spectral field distribution of the emitted pulse is recorded via a power monitor transversal to the nanowire axis. At the wavelength showing maximum power transmitted through the monitor, the spatial field distribution is exported.

\section{Results and discussion}
\label{ResultsSection}

\subsection{Lasing of nanowires on a dielectric substrate}
\label{silica_base}
The lasing model described above was first applied to a nanowire placed on top of a silica substrate.  In Sec. 3.3., these results were compared with those obtained considering the nanowire to be placed on metallic substrates, separated by a thin dielectric spacer.

\begin{figure}
\centering
    \includegraphics[width=1\textwidth]{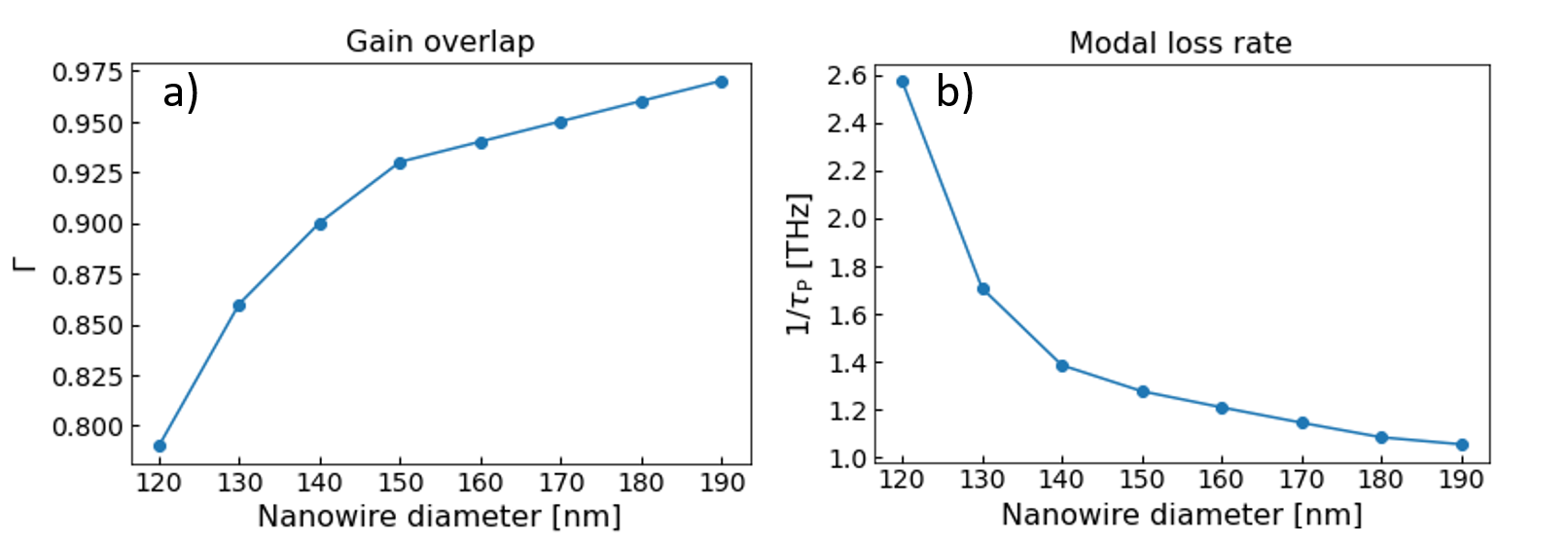}
\caption{a) Gain overlap and b) modal loss rate for the photonic mode of a ZnO nanowire placed on a silica substrate as a function of the nanowire diameter.}
\label{ref_mode}
\end{figure}

\paragraph{Mode characterization:}
The nanowire diameter is varied between $\SI{120}{\nano\meter}$ and $\SI{190}{\nano\meter}$ in steps of $\SI{10}{\nano\meter}$ to characterize size-dependent effects. This diameter range was chosen because the photonic mode is expected to be mainly responsible for lasing in this regime for a silica substrate, since the modes shown in \ref{Modes}b,c cannot exist because of the lack of plasmonic confinement. It also includes the range at which the photonic mode loses support via leakage into free space and thereby is interesting in terms of mode competition with plasmonic modes, which lack this cut-off. Since for diameters below $\SI{120}{\nano\meter}$, no nanowire-bound modes could be found, the analysis starts at $\SI{120}{\nano\meter}$. 
As can be seen in Fig. \ref{ref_mode}, the overlap with the gain material decreases with decreasing diameter, as the mode loses confinement at small diameters. The modal loss rate, on the other hand, increases, as the end facet reflectivity decreases as a consequence of mode leakage into the environment and free space.

\begin{figure}
\centering
    \includegraphics[width=1\textwidth]{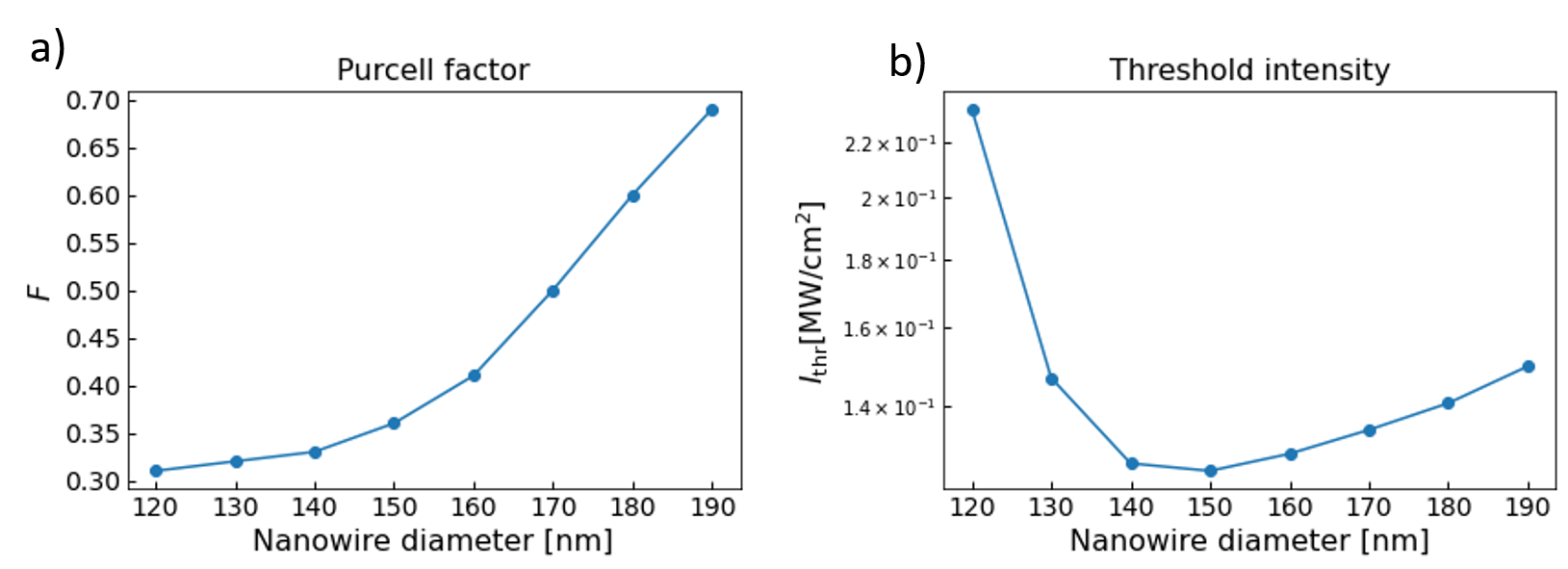}
\caption{a) Purcell factor for a dipole decaying at the nanowire center as a function of the nanowire diameter. b) Threshold intensity as a function of nanowire diameter.}
\label{PurcellTh_base}
\end{figure}

\paragraph{Purcell factor:}
The results of the Purcell factor calculation as described in Sec. \ref{NumericalMethods_Purcell} can be found in Fig. \ref{PurcellTh_base}a. The dipole was placed in the nanowire center, where the maximum of the photonic mode is located. The Purcell factor of the photonic mode drops with decreasing diameter as the mode increasingly leaks into the environment, thereby increasing the mode area. 

\paragraph{Lasing Threshold calculation:}
The threshold intensity, which is plotted in Fig. \ref{PurcellTh_base}b, depends strongly on the nanowire diameter. The threshold first decreases with decreasing diameter as a result of the fact that with smaller nanowires, less material loss needs to be compensated. The threshold reaches a minimum around 150 nm and increases again for further decreasing diameters since the photonic mode increasingly leaks into the surrounding medium. \ \\

\subsection{Dipole position for metallic substrates}
\label{pos_dipole}
\begin{figure}
\begin{subfigure}{0.5\linewidth}
  \centering
    \begin{overpic}[width=\textwidth]{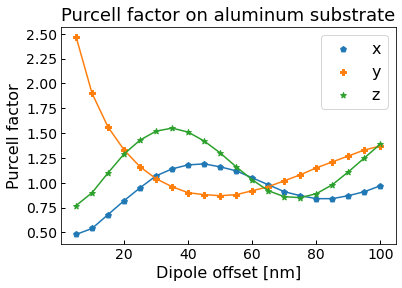}
        \put(45,95){a)}
      \end{overpic}
  \label{Thresholds_base_n}
\end{subfigure}
\begin{subfigure}{0.5\linewidth}
  \centering
      \begin{overpic}[width=\textwidth]{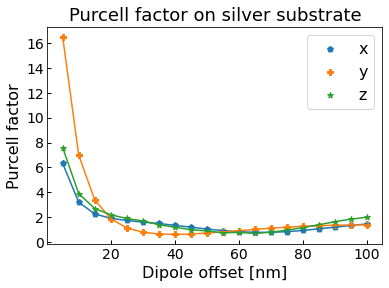}
        \put(45,95){b)}
      \end{overpic}
  \label{Thresholds_base_I}
\end{subfigure}
\caption{Purcell factor as a function of dipole position relative to nanowire-spacer interface (offset) for an a) aluminum and a b) silver substrate.}
\label{Purcell_position_sweep}
\end{figure}

Turning towards metallic substrates, aluminum and silver are identified as the most interesting substrates as they feature plasma frequencies in the UV spectral range. While the former has a plasma frequency at $\SI{80}{\nano\meter}$, the latter has a plasma frequency of $\SI{330}{\nano\meter}$. One would therefore expect a much stronger plasmonic enhancement for silver, since its plasma frequency is closer to the emission wavelength ($\lambda = \SI{380}{\nano\metre}$) of ZnO.\ \\
An increase of the Purcell factor when moving the dipole closer to the metal is expected, as the interaction with SPPs, which are exponentially confined to the interface between metal and dielectric, becomes stronger. In other words, the SPPs add a contribution to the LDOS only accessible for emitters close to the metal. The Purcell factor was analyzed as a function of position and orientation for the two metals for a nanowire with a diameter of 200 nm. Choosing a large diameter has the advantages of small mode leakage into free space and a large spatial separation between the maxima of the considered modes. The dipole position is parameterized in terms of the "Dipole offset", defined as the distance between dipole and the top of the dielectric spacer. The requirement of having at least one mesh cell between the dipole box walls and the emitter as well as having no material change inside the box leads to the minimum achievable distance between spacer and dipole of $\SI{5}{\nano\meter}$, given a uniform mesh with a 1 nm discretization.\ \\
Figure  \ref{Purcell_position_sweep} shows an increase of the Purcell factors for a polarization perpendicular to the metal surface (here $y$) when approaching the aluminum substrate and for all polarizations on the silver substrate. The $y$-polarization is dominant with respect to the plasmonic enhancement because plasmons are mainly polarized in the $y$-direction.\ \\
The fact that the Purcell factor increases for all polarization directions on silver as compared to only one on aluminum might be related to a closer match to the plasma frequency for silver, leading to a larger degree of confinement in the metal plane for the silver SPPs. \ \\
In the following, when calculating the Purcell factors of the plasmonic as well as the hybrid mode, the dipole is placed as close as possible to the metal. To be more precise, the dipole is placed 5 nm above the spacer-nanowire interface. Within numerical accuracy, this corresponds to the maximum of the plasmonic and hybrid mode within the nanowire. For calculating the Purcell factor of the photonic mode, the dipole is still put in the nanowire center.

\subsection{Lasing of nanowires on plasmonic substrates}

\subsubsection{Aluminum substrate}

\begin{figure}
\begin{subfigure}{0.5\linewidth}
  \centering
    \begin{overpic}[width=\textwidth]{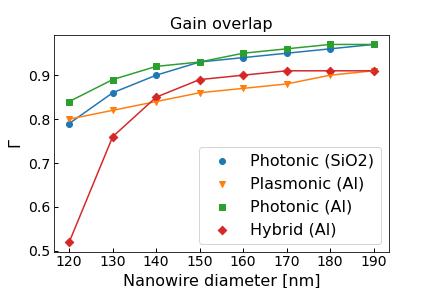}
        \put(25,85){a)}
      \end{overpic}
  \label{GainOverlap_base}
\end{subfigure}
\begin{subfigure}{0.5\linewidth}
  \centering
      \begin{overpic}[width=\textwidth]{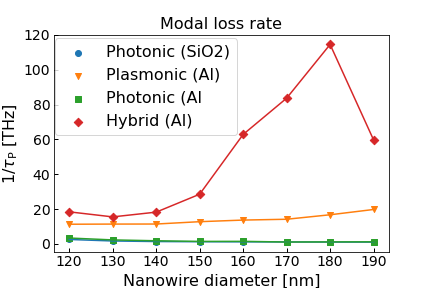}
        \put(22,85){b)}
      \end{overpic}
  \label{gp_base}
\end{subfigure}
\caption{a) Gain overlap and b) modal loss rate for the three modes of a nanowire placed on an aluminum substrate and the mode on a silica substrate as a function of nanowire diameter.}
\label{mode_calc_Al}
\end{figure}

\paragraph{Mode characterization:}
The gain overlap $\Gamma$ and the modal loss rate $1/\tau_{p}$ were analyzed, as can be seen in Fig. \ref{mode_calc_Al} and compared with the case of a silica substrate. The gain overlap (Fig. \ref{mode_calc_Al}a) is reduced for the plasmonic mode, but it is still significant. This is due to the high permittivity of ZnO, which is used for "weighting" the field distribution (see Eq. \ref{confinementA}). The photonic mode on aluminum shows a slightly higher gain overlap than the one on silica, since leakage into the substrate is reduced.\ \\
For large diameters, the hybrid mode has overlap values between the photonic and the plasmonic mode, as is expected, given its hybrid character. For smaller diameters, the hybrid mode cuts-off faster than the photonic mode, and therefore has a reduced gain overlap.\ \\
The modal loss rate (Fig. \ref{mode_calc_Al}b) of the plasmonic mode and the hybrid mode are increased strongly with respect to the photonic modes, which is mainly a result of the non-negligible propagation losses, and decreases with decreasing diameter. Interestingly, the modal loss rate of the hybrid mode reaches a value as high as ca. 115 THz at about 180 nm. This is caused by high propagation losses and a high modal group velocity around this point (see supplementary document Sec. 2C).

\begin{figure}
\centering
    \includegraphics[width=1\textwidth]{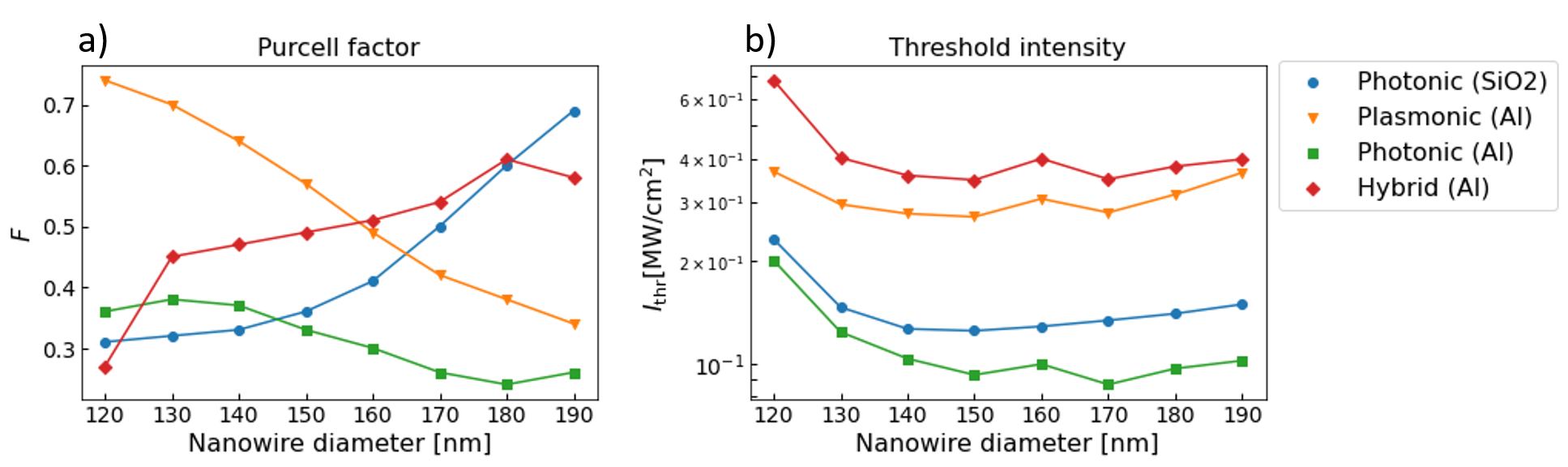}
\label{Purcell_Al}
\caption{a) Purcell factor for a dipole decaying at the nanowire bottom (plasmonic \& hybrid modes) and the nanowire center (photonic modes for both substrates) as a function of the nanowire diameter. b) Threshold intensity of the modes under consideration as a function of nanowire diameter.}
\label{Purcell_Th_Al}
\end{figure}

\paragraph{Purcell factor:}
Another influence of the aluminum substrate is a modification of the Purcell factor. To characterize the Purcell factors of the three modes, in separate simulations, one dipole is placed in the center of the nanowire and one at the bottom. The Purcell factors $F_{x,y,z}$ were calculated and averaged as in Sec. \ref{silica_base}. In Fig. \ref{Purcell_Th_Al}a, it is shown that the Purcell factor of the photonic mode is reduced for the aluminum substrate, which supports more modes that compete for gain.\ \\
Interestingly, the hybrid mode as well as the plasmonic mode have an increased Purcell factor. But while the hybrid mode dominates at large diameters, the plasmonic mode dominates at small diameters, whereas the hybrid mode cuts off at small diameters.\ \\

\paragraph{Lasing Threshold calculation:}
The threshold intensities of all three modes were calculated and compared with the threshold intensity on a silica substrate in Fig. \ref{Purcell_Th_Al}b. The plasmonic mode and the hybrid mode on the aluminum substrate have the highest thresholds. Their threshold depends on the nanowire diameter similarly as the photonic mode on a silica substrate, as expected from their weak degree of confinement to the spacer. \ \\
The photonic modes have similar thresholds in the small diameter range independently of the substrate. Remarkably, for larger diameters, the aluminum substrate leads to a threshold reduction, since leakage into the substrate is reduced, but effectively no coupling to highly dissipative SPPs is taking place.\ \\
We conclude that on aluminum, the photonic mode dominates the lasing process under steady-state conditions. The losses of the plasmonic and the hybrid mode cannot be compensated by the increased Purcell factor. \ \\

\subsubsection{Silver substrate}
\label{silver}
In contrast to the case of the aluminum substrate, the high losses of silver in the UV spectral range of interest prevent the possibility of pure plasmonic lasing with field distributions as shown in Fig. \ref{Modes}b. For example, for a nanowire diameter of $\SI{120}{\nano\meter}$, the photonic mode effective refractive index has an imaginary part of 0.009 and the hybrid mode 0.3, whereas the plasmonic mode effective refractive index has an imaginary part of 1.67. Therefore, we only consider the photonic and the hybrid modes. Note, however, that hybrid modes, as sketched in Fig. \ref{Modes}c, also lead to the excitation of plasmons, but suffer smaller losses.\ \\

\begin{figure}
  \centering
\includegraphics[width=\textwidth]{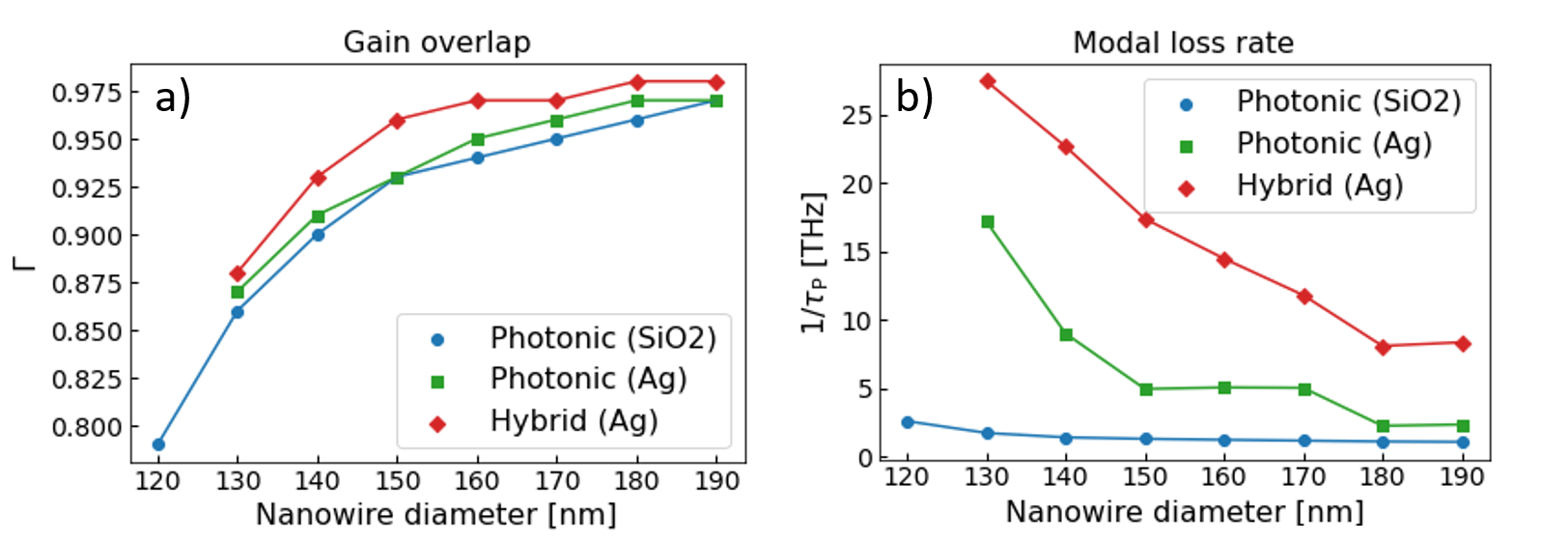}
\caption{a) Gain overlap and b) modal loss rate for the two modes of a nanowire placed on a silver substrate and the mode on a silica substrate as a function of the nanowire diameter.}
\label{mode_calc_Ag}
\end{figure}

\paragraph{Mode characterization:}
The modes could only be clearly distinguished starting at $\SI{130}{\nano\metre}$. The photonic modes have a similar overlap with the gain material, as can be seen in Fig. \ref{mode_calc_Ag}a. The hybrid mode, on the other hand, has an increased overlap with the gain medium caused by the field enhancement close to the metal. The modal loss rates (Fig. 8b) for both modes are increased significantly compared to the photonic mode on a silica substrate. Furthermore, the losses of both modes increase with decreasing diameter because the mode fraction overlapping with the gain material leaks into free space.

\begin{figure}
\centering
    \includegraphics[width=1\linewidth]{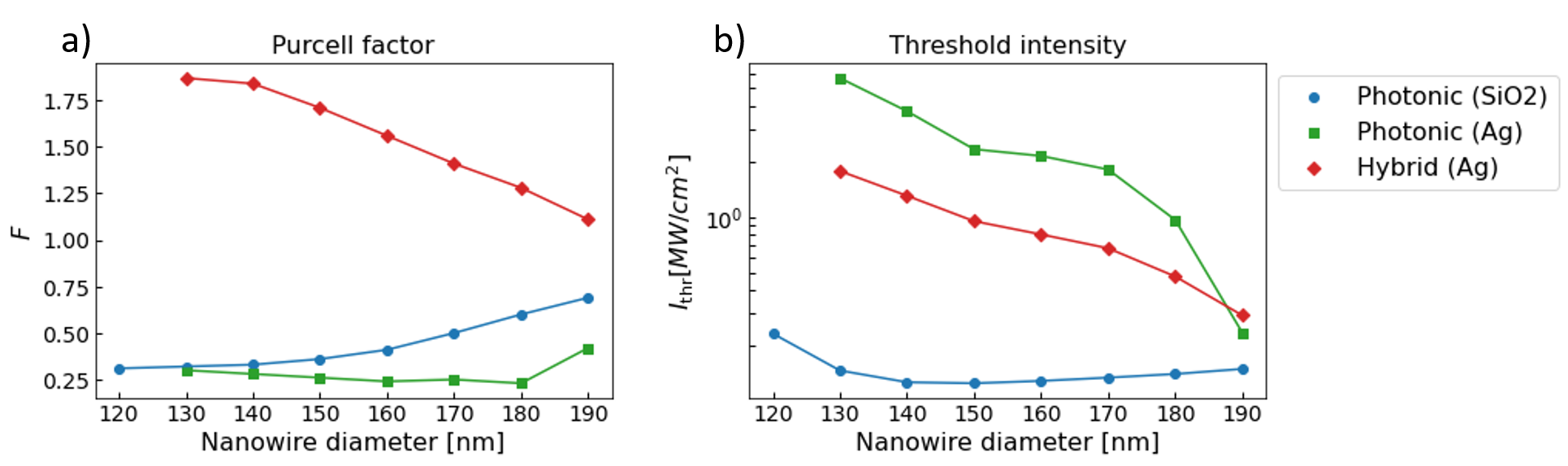}
\caption{a) Purcell factor for a dipole decaying at the nanowire bottom (hybrid mode) and the nanowire center (photonic modes for both substrates) as a function of the nanowire diameter for a silver substrate. b) Threshold intensities of the modes under consideration as a function of nanowire diameter.}
\label{PurcellTh_Ag}
\end{figure}

\paragraph{Purcell factor calculation:}
The Purcell factors are much smaller than expected from Fig. \ref{Purcell_position_sweep}b, since most of the high Purcell factor is attributed to the highly damped plasmonic mode. The Purcell factor of the hybrid mode is small for high diameters, but increases at lower diameters, as can be seen in Fig. \ref{PurcellTh_Ag}a. 

\paragraph{Lasing Threshold calculation:}
The threshold intensities plotted in Fig. \ref{PurcellTh_Ag}b show a large increase overall and with respect to the nanowire diameter. For almost all but very high diameters, the hybrid mode will dominate the lasing process, since the increases in gain overlap and Purcell factor can compensate the higher propagation losses observed with respect to the photonic mode on aluminum. \ \\

\subsubsection{FDTD calculations of field distribution at the lasing wavelength}
\begin{figure}
\centering
    \includegraphics[width=1\linewidth]{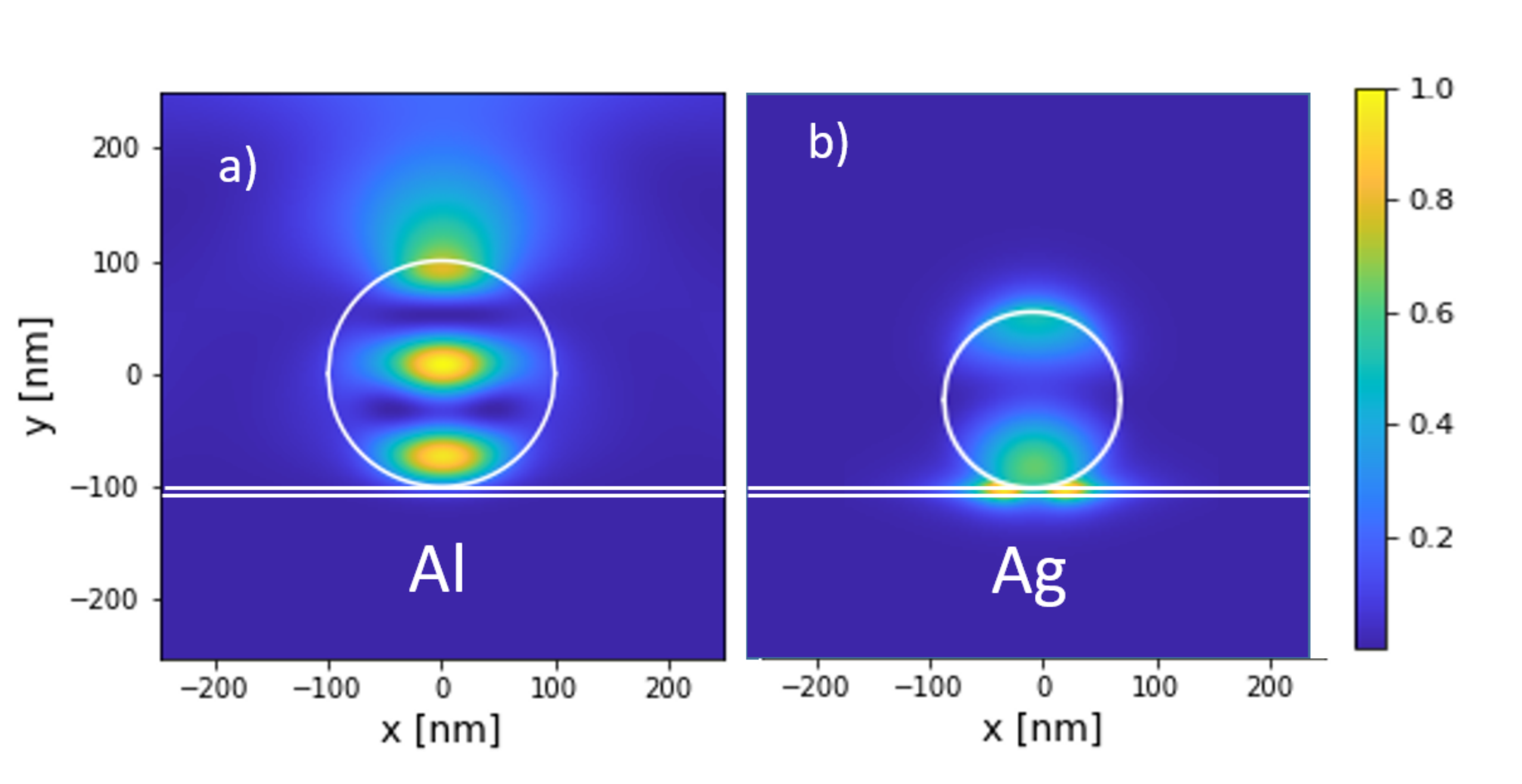}
\caption{Distribution of $|E|^2$, normalized to the maximum value, in the lasing regime for a ZnO nanowire of diameter a) 200 nm on top of an aluminum substrate and b) 160 nm on top of a silver substrate. A 5 nm silica spacer was included in both cases. The nanowire outline as well as the spacer are indicated in white.}
\label{FDTD_modes}
\end{figure}

As explained in Sec. \ref{FDTD}, FDTD calculations were performed to corroborate the findings presented above. The results of these FDTD calculations for two nanowires with comparable diameters and excitation conditions but two different substrates are presented in Fig. \ref{FDTD_modes}.

Evidently, the field distribution for the case of an aluminum substrate is concentrated inside the nanowire, whereas for a silver substrate, excitation of SPPs close to the metal surface can clearly be observed, while still a strong overlap with the nanowire is present.\ \\
The field distributions look different to the ones shown in Fig. \ref{Modes}. Figure \ref{FDTD_modes}a shows three maxima, whereas \ref{Modes}a only shows one. The reason is that our choice of seeding the lasing modes via shutting of the pump excites a set of modes that is polarized perpendicular to the waveguide mode in \ref{Modes}a. Nonetheless, the calculations show that very little interaction with surface plasmons takes place.\ \\
The mode shown in Fig. \ref{FDTD_modes}b, on the other hand, more closely resembles \ref{Modes}c because of the closeness of the lasing wavelength to the plasma frequency. The main difference lies in the fact that the plasmonic part in \ref{FDTD_modes}b shows 2 lobes, whereas \ref{Modes}a only shows one. This is due to a slight difference in the fitting of the material properties between both configurations, which has a large influence on the field distribution.\ \\
Together, both calculations show that lasing for a silver substrate is dominated by hybrid modes, whereas for aluminum, little interaction with surface plasmons is taking place.

\section{Conclusions}
We have set out to quantify the effect of metallic substrate materials on the lasing thresholds of ZnO nanowire lasers. In the process, we have derived an intuitive method to quantify the effect of plasmonically-induced Purcell factor enhancement on silver and aluminum substrates. We presented detailed mode calculations and a rate-equation model to study  mode competition.\ \\
Considering silver, a plasmonic material with a plasma frequency being resonant with the bandgap of ZnO, we have shown that plasmonic lasing is a double-edged sword: On the one hand, the low mode volume of SPP-modes offers interesting possibilities with respect to the strong increase of the Purcell factor. On the other hand, the high  propagation losses make an excitation of these modes in a steady-state or quasi-steady-state regime very difficult.\ \\
Note that these results offer a new perspective for the interpretation of previously published results \cite{Oulton2009}. A careful analysis of the mode profiles shows that the modes found to be responsible for lasing by Oulton et. al. on the plasmonic substrate are in fact the "hybrid" modes in our terminology. The order-of-magnitude increase below a 100 nm nanowire diameter fits nicely with the cut-off expected for nanowire-bound modes (at around 120 nm diameter, the photonic modes cannot be found in our solver) and the losses of the purely plasmonic modes (see introductory text for Sec. \ref{silver}).\ \\
As expected, silver shows a strong effect on lasing properties and even offers the possibility to access hybrid modes in the steady-state by compensating increased losses by increased Purcell factors and gain overlap. Aluminum, on the other hand, does not offer these advantages. But curiously, an aluminum substrate decreases the threshold for photonic lasing as compared to a photonic substrate, since the small penetration depth suppresses mode leakage into the substrate.\ \\
These findings should also be valid for other combinations of nanowire materials and metals. We expect for metals with plasma frequenciess far below the nanowire emission wavelength $\lambda$ and with diameters comparable to $\lambda/n$, where $n$ is the nanowire refractive index at the emission wavelength, a threshold reduction without plasmon excitation. This is caused by a reduction of leakage into the substrate or, in other words, a small penetration depth into the metal. On the other hand, for resonant combinations, we expect a threshold increase caused by plasmon excitation via Purcell enhancement.\ \\
Another interesting finding is related to the polarization properties of emitters close to the substrate. Highly-resonant combinations of nanowire lasing wavelength and plasma frequency seem to offer an increase of the emission rate independently of the emitter polarization, whereas in an off-resonant configuration, mainly emitters polarized normal to the metal experience enhancement.

\section*{Acknowledgement}
We acknowledge the financial support via the project ID 398816777 in the CRC 1375 “NOA–Nonlinear optics down to atomic scales” funded by the Deutsche Forschungsgemeinschaft (DFG).\ \\ 
A.B.  thanks the financial support of the German Research Foundation (DFG) under the framework of the International Research Training Group 
(IRTG) 2675 "META-ACTIVE".

\printbibliography

\end{document}


\maketitle

\section{Theoretical Model}

\subsection{Purcell factor}
Generally, the emission problem can be solved via Green's function methods. The Purcell factor can be quantitatively connected to it, as is derived in this section.\ \\
$\overleftrightarrow{G}(\textbf{r}, \textbf{r'})$ in electrodynamics is defined by \cite{NovotnyHecht}

\begin{equation}
    \nabla\times\nabla\times\overleftrightarrow{G}(\textbf{r}, \textbf{r'},\omega) - k^2 \overleftrightarrow{G}(\textbf{r}, \textbf{r'},\omega) = \overleftrightarrow{I} \delta(\textbf{r}-\textbf{r'}),
    \label{GreenDefinition}
\end{equation}
where $k$ is the wavenumber of radiation, $\omega$ is the angular frequency and $\overleftrightarrow{I}$ is the identity matrix.\ \\
The Green's function is directly related to the decay rate of a two-level quantum system $\Gamma$ via Fermi's Golden rule expressed as (cf. \cite{OpticalAntennas}):
\begin{equation}
    \Gamma = \frac{\pi\omega}{3\hbar\epsilon_0} |\langle g|\hat{\textbf{p}}|e\rangle|^2 \rho_\textbf{p}(\textbf{r},\omega).
    \label{FermiRule}
\end{equation}
Here, $\omega$ is the transition frequency, $|\langle g|\hat{\textbf{p}}|e\rangle|$ is the transitional dipole element between the ground and excited states and $\rho_\textbf{p}(\textbf{r},\omega)$ is the local density of optical states (LDOS) at frequency $\omega$ and position \textbf{r}, which can further be expressed as:
\begin{equation}
    \rho_\textbf{p}(\textbf{r},\omega) = \frac{6\omega}{\pi c^2}[\textbf{n}_p Im(\overleftrightarrow{G}(\textbf{r}, \textbf{r},\omega)) \textbf{n}_p],
    \label{LDOS}
\end{equation}
with $\textbf{n}_p$ the normal vector in the direction of the dipole.\ \\
Taking into account that $\rho_\textbf{p}$ in Eq. \ref{FermiRule} for free space can be expressed as $\rho_0 = \omega^2/(\pi^2c^3)$ \cite{OpticalAntennas}, the Purcell factor can be defined as:

\begin{equation}
    F(\textbf{r}) = \frac{\Gamma}{\Gamma_0} = \frac{\pi^2 c^3}{\omega^2} \rho_\textbf{p}(\textbf{r},\omega).
    \label{F_and_LDOS}
\end{equation}

Further analysis shows that the LDOS can be given by

\begin{equation}
    \rho_\textbf{p}(\textbf{r},\omega)=\frac{\omega^2}{\pi^2 c^3} \frac{P(\textbf{r})}{P_0(\textbf{r})},
\end{equation}
with $P$ and $P_0$ the powers radiated by a classical dipole in the structure under investigation and in free space at position \textbf{r}, respectively. This leads to a simple expression for the determination of the Purcell factor in an arbitrary structure \cite{OpticalAntennas}:

\begin{equation}
    F= \frac{P}{P_0} = \frac{P_{\text{rad}} + P_{\text{nrad}}}{P_0},
    \label{Purcell_factor_radNonrad}
\end{equation}
where $P_{\text{rad}}$ and $P_{\text{nrad}}$ are the radiative and non-radiative emitted power in the presence of the structure by a classical dipole, respectively. $P_0$ is the power emitted by the same dipole placed in free space.  \ \\
The calculation of the Purcell factor by means of Eq. \ref{Purcell_factor_radNonrad} can be performed by different numerical methods like Finite Element Method (FEM) or Finite-Difference Time-Domain (FDTD) methods. In particular, this last method was successfully used for obtaining the spontaneous emission rate of microdisk cavities \cite{Yariv1} and photonic crystals \cite{PC1}. For a rigorous, quantum mechanical model of fluorescence, we refer to Refs. \cite{Busch} and  \cite{Yariv2}.\ \\

\section{Simulation results}
\subsection{Propagation losses}
\begin{figure}
\begin{subfigure}{0.5\linewidth}
  \centering
    \begin{overpic}[width=\textwidth]{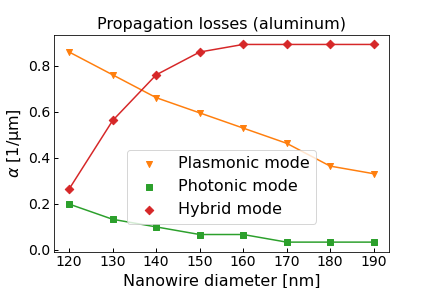}
        \put(25,85){a)}
      \end{overpic}
  \label{GainOverlap_base}
\end{subfigure}
\begin{subfigure}{0.5\linewidth}
  \centering
      \begin{overpic}[width=\textwidth]{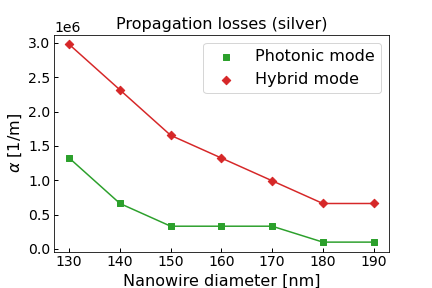}
        \put(25,85){b)}
      \end{overpic}
  \label{gp_base}
\end{subfigure}
\caption{Propagation losses of the investigated modes for an a) aluminun and a b) silver substrate.}
\label{Alphas}
\end{figure}

The propagation losses $\alpha$ are related to the imaginary part of the effective refractive mode index $\kappa$ via 

\begin{equation}
    \alpha = 2\cdot 2\pi \cdot \frac{\kappa}{\lambda}.
\end{equation}
Notice that the contribution to the modal loss rate is calculated by multiplying $\alpha$ by the modal group velocity. The values of the propagation losses for the two wire materials are shown in Fig. \ref{Alphas}.\ \\
As expected, the modes showing plasmonic contributions have increased propagation losses compared to the photonic modes. Almost all losses increase when decreasing the nanowire diameter, since the modes increase leak into free space. The hybrid mode on aluminum, however, has reduced propagation losses at low diameters. We believe this to be caused by the fact that the hybrid nature changes with diamater - the mode is more dominated by its plasmonic part at large diameters, but becomes increasingly dominated by its "photonic" contribution at small diameters because of the cut-off. This behavior could also be observed for the Purcell factors (see main text).\ \\

\subsection{End facet reflectivity}
To calculate the endfacet reflectivity for specific modes, we perform an additional FDTD calculation for every mode. This calculation consists of extending only one end of the nanowire into the PML. Then, we excite a specific nanowire mode via a mode source $\SI{750}{\nano\meter}$ away from the free standing facet and with the propagation vector pointing towards the free end. \ \\
As the mode reaches the endfacet, part of it gets reflected and part of it gets transmitted. We monitor the transmitted power $\SI{50}{\nano\meter}$ away from the endfacet and call this the transmitted power $T$, normalized to the injected source power. The reflected power $R$ gets completely absorpted by the PML. Since we know the imaginary part of the effective refractive index (see Sec. 2.1) of the mode, we can calculate the propagation losses $A$ and calculate the reflectivity $R$ via

\begin{equation}
    R = 1-A-T.
\end{equation}

\begin{figure}
\begin{subfigure}{1\linewidth}
  \centering
    \begin{overpic}[width=0.8\textwidth]{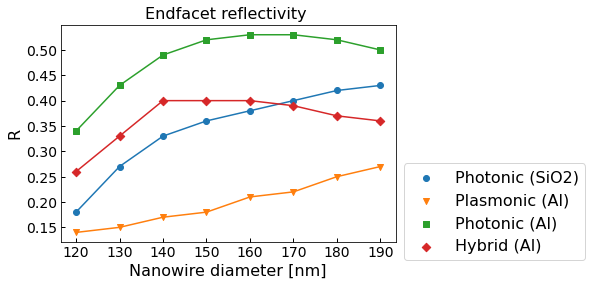}
        \put(35,100){a)}
      \end{overpic}
  \label{GainOverlap_base}
\end{subfigure}
\begin{subfigure}{1\linewidth}
  \centering
      \begin{overpic}[width=0.8\textwidth]{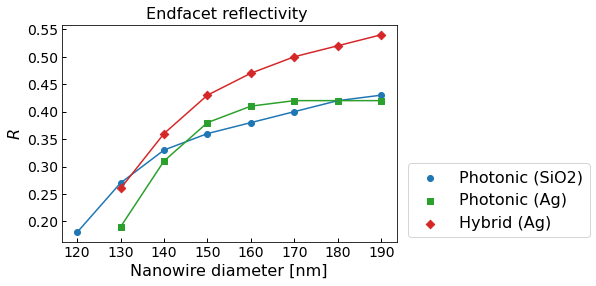}
        \put(35,100){b)}
      \end{overpic}
  \label{gp_base}
\end{subfigure}
\caption{Endfacet reflectivities of the investigated modes for an a) aluminun and a b) silver substrate.}
\label{reflectivities}
\end{figure}

Calculations of the endfacet reflectivity can be seen in Fig. \ref{reflectivities}

In the aluminum substrate, the endfacet reflectivity of the photonic mode is increased compared to a silica substrate because of reduced leakage into the substrate, whereas the plasmonic mode has a reduced reflectivity because a larger part of the mode intensity (as compared to the photonic mode) is not located inside the nanowire. The hybrid mode has a reflectiviy similar to the photonic mode on the dielectric substrate.\ \\
For silver, the photonic modes show similar reflectivity values, while the hybrid mode's reflectivity is slightly enhanced. We attribute this increase to the higher overlap with the gain material caused by the field enhancement close to the metal.

\subsection{Modal group velocity}
\begin{figure}
\begin{subfigure}{0.5\linewidth}
  \centering
    \begin{overpic}[width=\textwidth]{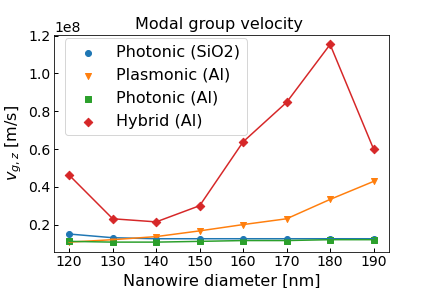}
        \put(25,85){a)}
      \end{overpic}
  \label{GainOverlap_base}
\end{subfigure}
\begin{subfigure}{0.5\linewidth}
  \centering
      \begin{overpic}[width=\textwidth]{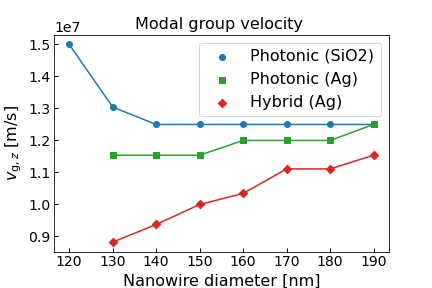}
        \put(35,85){b)}
      \end{overpic}
  \label{gp_base}
\end{subfigure}
\caption{Group velocities of the investigated modes for an a) aluminun and a b) silver substrate.}
\label{Group_velocities}
\end{figure}

The modal group velocity is calculated via the group index, given by

\begin{equation}
    \dv{(\omega n)}{\omega} \approx \frac{n_2\omega_2-n_1 \omega_1}{\omega_2 - \omega_1} = \frac{\frac{n_2}{\lambda_2}-\frac{n_1}{\lambda_1}}{\frac{1}{\lambda_2} - \frac{1}{\lambda_1}},
\end{equation}
where the $i$ refers to different free-space wavelengths and $\lambda_i$ are the free-space wavelengths of the calculates modes. The dispersion of all materials, including ZnO, is taken into account. The group velocity plots are presented in Fig. \ref{Group_velocities}.

For both metals, the photonic modes have similar group velocities. The plasmonic mode and the hybrid mode on aluminum, however, have a much higher group velocity for larger wires, since the overlap with the highly dispersive wire material is smaller (dispersive refering to the refractive index). We have no clear understanding about why the hybrid mode on aluminum shows this strongly increased mode velocity, but it does not affect the main findings of the article, namely that on an aluminum substrate, the lasing threshold can be reduced. The hybrid mode on silver has an increased overlap with the strongly dispersive nanowire, which leads to an increase of the mode index, thereby reducing the group velocity. Furthermore, the high group velocity of the hybrid mode on aluminum is caused by a small degree of confinement.\ \\

\subsection{Threshold errors}
\begin{figure}
\begin{subfigure}{0.5\linewidth}
  \centering
    \begin{overpic}[width=\textwidth]{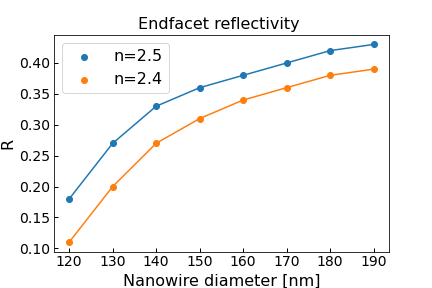}
        \put(25,65){a)}
      \end{overpic}
  \label{GainOverlap_base}
\end{subfigure}
\begin{subfigure}{0.5\linewidth}
  \centering
      \begin{overpic}[width=\textwidth]{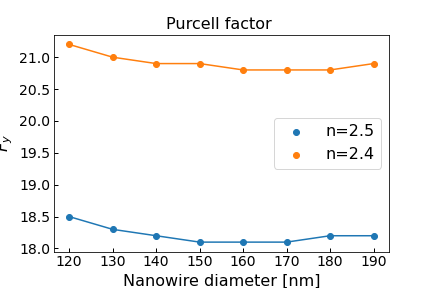}
        \put(25,85){b)}
      \end{overpic}
  \label{gp_base}
\end{subfigure}
\caption{a) Endfacet reflectivity for a nanowire on a photonic substrate and b) Purcell factors for a dipole oriented along $y$ in a nanowire placed on a silver substrate (indcluding the spacer) for two different refractive indices of the nanowire.}
\label{Corrections}
\end{figure}

As has been said in the main text, the chosen value of $n=2.5$ for the nanowire material may result in errors concerning the threshold. To quantify the effect, we calculated the endfacet reflectivity for a nanowire on top of a silica substrate as well as the Purcell factor for a dipole positioned close to the silver substrate also for a refractive index of $n=2.4$, as shown in Fig. \ref{Corrections}. The differences represent corrections to the main result, but do not alter the qualitative statements of the main text regarding which modes dominate the lasing process.

\FloatBarrier
\printbibliography